\documentclass[preprint,eqsecnum,aps]{revtex4}

\usepackage{graphicx}
\usepackage{dcolumn}
\usepackage{amsmath}

\begin{document}

\title{Initial Parton Distribution just after Heavy Ion Collisions}

\author{Ghi R. Shin$^{(1)}$ and Kang S. Lee$^{(2)}$}
\affiliation{ $^{1)}$Department of Physics, Andong National University,
Andong, South Korea\\
$^{3)}$Department of Physics, Chonnam National University, Kwangju, South Korea}

\date{\today}
~ \\

\begin{abstract}
We study the initial distribution of a parton system which is formed just after
relativistic heavy ion collision by the elastic scattering among the constituent partons in details
and analyze the baryon and strangeness contents of the primary parton system.
We present the rapidity and energy distributions of the system.
\end{abstract}

\maketitle
\newpage
\section{Introduction}
The relativistic heavy ion collision at RHIC or LHC energy liberates lots of constituent partons
of nuclei by primary collisions and those partons evolve according to quantum theory 
and may become a quark-gluon plasma which has been searched intensively by AGS and RHIC.
The study of the early system of those partons thus are very important and
we somehow have to know the momentum distribution as well as the space distribution
to understand in details the dynamics of the system as a function of time.

There are several schemes to obtain the initial (phase space) distribution of partons: 
The constituent partons of a nucleus can be characterized by the Feynman x where the soft parton
has small x and the hard one large x. The soft parton can be identified as 
the non-abelian Weizsacker-Williams field radiated from the hard partons and is dominant in number. 
While the two nuclei pass through each other in collision, the high x partons of one nucleus can be scattered
off from those high x partons of the other nucleus and the non-abelian Weicsacker-Williams field
lose the coherence of the parent parton and become particles. This initial parton information has been
studied by Krasnitz et al\cite{knv03} and Lappi\cite{lap03} based on CGC\cite{mcl94,am00}, namely, 
shattering of the color glass condensates of two colliding nuclei, and
the analytic distribution function has been proposed. It however can be applicable to the central rapidity
region only and the production of quarks or antiquarks can not be calculated from the scheme.

Another theory to produce dynamically the partons after a heavy ion collision is that while the colliding nuclei
is passing through each other, they rapidly decelerate and produce Schwinger-like radiation already in thermal
spectrum\cite{kha06}. The background field are the pulse-like CGC of McLerran-Venugopalan model of the other
nucleus. This theory could provide the scenario to explain the fast thermalization of the parton system. 

Even though we can understand the gluon production from these studies, those do not provide the initial baryon density
which is shown in RHIC experiments\cite{star05,phen05}. To understand this phenomena we certainly
have to include quarks as well as antiquarks from the primary collision.
Consider the CM frame of two colliding nuclei at high energy.
The colliding nuclei are Lorentz contracted to become disks and the transversal motion of constituent partons
can be ignored  since the longitudinal momentum is dominant at high energy. 
When the colliding nuclei pass through each other, the constituents of a projectile nucleus make scattering
with those of a target nucleus(note that the partons in the same nucleus do not collide with each other
since the time delation) and become particles if the momentum transfer is large enough.
This, so-called, factorization method, has been used to calculate jets or minijets\cite{eks96,nayak,ham99,coo02}
from early 80s. There are many schemes to obtain the parton distribution of a hadron or a nucleus\cite{szc}: 
for examples, we can fold the parton distribution of a nucleon and the nucleon distribution
of a nucleus\cite{grv98,eks99}.
The wave function of a high energy nucleus has been studied extensively and we use GRV98, CTEQ and 
an emperical distribution in our study.
Since this method only gives relatively high $p_\perp$ partons but soft partons can not be obtained so that
it gives a reasonable distribution at higher energy but does not give good at RHIC energy.
The studies on this direction had been persuited by Eskola\cite{eks96}, by Hammon et al.\cite{ham99},
by Nayak et al.\cite{nayak} and by Cooper et al.\cite{coo02}.
We however think it is worth to extend the study to include baryon and strangeness contents at RHIC
and LHC energy. We also briefly note here that the spatial distribution of produced partons can be understood
assuming that the quanta becomes particle after the time $1/E_T$, which $E_T$ is the transversal energy of the quanta
and at the place where the quanta travels ${\vec p} /EE_T $ from the scattering point.

We rewrite the basic formulas to calculate the production of partons in Section II. 
We present the numerical results in Section III. We summarize and conclude in Section IV.

\section{Initial parton production}

We assume that partons are produced at relativistic heavy-ion collisions
by elastic scattering between the constituent of a projectile nucleus and that of a target nucleus. 
Since the scattering cross section gives the number of collision events per unit overlap area,
we can write the total number of events \cite{eks96,ham99,coo02,book}:
\begin{eqnarray}
{N^{event}} &=& K T(b) \int dy_3 dy_4 d^2 p_T 
\sum_{ij,\,kl} x_1 f_{i/A}(x_1,Q_0^2) x_2 f_{j/B}(x_2,Q_0^2)
{{d \sigma^{ij\rightarrow kl}(\hat{s},\hat{t},\hat{u})} \over
{d {\hat t}}},
\label{pt_y_dis}
\end{eqnarray}
where a parton $i$ of the nucleus $A$ makes a collision with a parton $j$ of the nucleus $B$ and produces
partons $k$ and $l$. Each parton has rapidity $y_1$, $y_2$, $y_3$ and $y_4$, respectively. $x_1$ and $x_2$ are
the Bjorken scaling variables of parton $i$ and $j$.
The relations between the variables before and after the collision are given by
\begin{eqnarray}
x_1 &=& p_T (e^{y_1} + e^{y_2}) /\sqrt{s},\\
x_2 &=& p_T (e^{-y_1} + e^{-y_2}) /\sqrt{s},\\
\hat{s} &=& x_1 x_2 s, \\
\hat{t} &=& -p_T^2 (1 +e^{y_2 - y_1}),\\
\hat{u} &=& -p_T^2 (1 +e^{y_1 - y_2}).
\end{eqnarray}
We also rewrite the available kinematic region for convenience \cite{ham99,shin02,shin03},
\begin{eqnarray}
Q_0\, ^2 \leq p_T\, ^2 \leq ({{\sqrt{s}}\over{2 \cosh y}})^2, \\
-\log ({{\sqrt{s}}\over{p_T}} - e^{-y}) \leq y_4 \leq 
\log({{\sqrt{s}}\over{p_T}} - e^{-y}),\\
|y| \leq \log ( {{\sqrt{s}}\over{2 Q_0}} + \sqrt{
{{s}\over{4Q_0\,^2}} - 1}).
\end{eqnarray}
$K$ is the K-factor to include the higher-order diagrams; we will set $K = 2$ throughout our study.
$Q_0$ is the momentum scale which we are looking at the nucleus and is a minimum momentum transfer.
The hat on the Mandelstan variables means that those are the variables of a parton.

$T(b)$ is the nuclear geometric factor at a given impact parameter $b$. Assuming that the density of nucleus
is constant over the sphere of radius $R$ with the sharp edge, we can write
\begin{eqnarray}
T( \vec b) &=& \int d \vec r T_{AB} (\vec r ;\vec b ) \\ \nonumber
&=& 4 \rho_0^A \rho_0^B \int d {\vec r} \sqrt{R_A^2-(\vec r - \vec b/2)^2}
\sqrt{R_B^2-(\vec r + \vec b/2)^2},
\label{overlap_fn}
\end{eqnarray}
where $T_{AB}$ is the nuclear thickness function. Note that we assume there
is no correlation between the position and the momentum of a given parton.

Since each event gives two partons, we can write the distribution of produced partons as follows
\begin{eqnarray}
{dN^{jet}}\over{dp_T dy} &=& K T(b) \int dy_4 {{2\pi p_T} \over{\hat s}}
\sum_{ij,\,kl} x_1 f_{i/A}(x_1,p_T^2) x_2 f_{j/B}(x_2,p_T^2)
{1 \over 2} {{d \sigma^{ij\rightarrow kl}(\hat{s},\hat{t},\hat{u})}\over{d{\hat t}}},
\label{pt_y_dis}
\end{eqnarray}
Note also that a quark minijet of $(y, p_T)$ from a process,  for example, 
$gq \rightarrow gq$, comes from the scattering of a quark of the nucleus A and a gluon of the nucleus B
via the t-channel or from the scattering of a gluon of the nucleus A and a quark of the nucleus B via the u-channel
so that we have to include both the $\sigma_{gq \rightarrow gq}(\hat t, \hat u)$ and the
$\sigma_{gq \rightarrow gq}(\hat u, \hat t)$ processes multiplied by an approperate distribution, respectively.

The processes we consider in our study are 
\begin{eqnarray}
gg &\leftrightarrow& gg, q \bar q,\\ 
g q &\leftrightarrow& gq,\\ 
g \bar q &\leftrightarrow& g \bar q, \\
q^a q^b &\leftrightarrow& q^c q^d , \\
q\bar q &\leftrightarrow& q \bar q,\\
\bar q^a \bar q^b &\leftrightarrow& \bar q^c \bar q^d.
\end{eqnarray}
We however ignore some of basic channels such as $qg \rightarrow q \gamma$, 
$q \bar q \rightarrow \gamma \gamma$, $q \bar q \rightarrow g \gamma$,
which could provide important information on the system.

The cross sections for the processes up to the leading order (LO) are
\begin{eqnarray}
{{d\sigma^{gg\rightarrow gg}}\over{d \hat t}} &=&
{{9\pi\alpha_s^2}\over{2{\hat s}^2}}(3-{{{\hat t}{\hat u}}\over {\hat s}^2}
- {{{\hat s}{\hat u}}\over{{\hat t}^2}}
- {{{\hat s}{\hat t}}\over{{\hat u}^2}} ) \\ \\		\label{gg_gg}
{{d\sigma^{gg\rightarrow q_a\bar q_b}}\over{d{\hat t}}} &=&
{{\pi\alpha_s^2}\over{6{\hat s}^2}} \delta_{ab} ( {{\hat u} \over {\hat t}} 
+ {{\hat t} \over {\hat u}} - {9\over 4}{{{\hat t}^2+{\hat u}^2}\over{{\hat s}^2}} ) 
\\ \\	\label{gg_qbq}
{{d\sigma^{gq\rightarrow gq}}\over{d{\hat t}}} &=&
{{4\pi\alpha_s^2}\over{9{\hat s}^2}} ( - {{\hat u}\over {\hat s}} - {{\hat s}\over {\hat u}} 
+ {9 \over 4}{{{\hat s}^2+{\hat u}^2}\over{{\hat t}^2}}) \\  \\	\label{gq_gq}
{{d\sigma^{q_aq_b\rightarrow q_aq_b}}\over{d{\hat t}}} &=&
{{4\pi\alpha_s^2}\over{9{\hat s}^2}}[{{{\hat s}^2+{\hat u}^2}\over{{\hat t}^2}}
+\delta_{ab}({{{\hat t}^2+{\hat s}^2}\over{{\hat u}^2}}
-{2\over 3}{{{\hat s}^2}\over{{\hat u}{\hat t}}})] \\  \\		\label{qq_qq}
{{d\sigma^{q_a\bar q_b\rightarrow q_c\bar q_d}}\over{d{\hat t}}} &=&
{{4\pi\alpha_s^2}\over{9{\hat s}^2}}[ \delta_{ac}\delta_{bd}
{{{\hat s}^2+{\hat u}^2}\over{{\hat t}^2}} + \delta_{ab}\delta_{cd}
{{{\hat t}^2+{\hat u}^2}\over{{\hat s}^2}}
-\delta_{abcd}{2\over 3}{{\hat u}^2\over{{\hat s}{\hat t}}}] \\ \\	\label{qbq_qbq}
{{d\sigma^{q_a\bar q_b\rightarrow gg}}\over{d{\hat t}}} &=&
{{32\pi\alpha_s^2}\over{27{\hat s}^2}} \delta_{ab} [{{\hat u} \over {\hat t}}
+{{\hat t}\over {\hat u}}-{9\over 4}{{{\hat t}^2+{\hat u}^2}\over{{\hat s}^2}}]
\label{qbq_gg}
\end{eqnarray}
and
\begin{eqnarray}
{{d\sigma^{g\bar q\rightarrow g\bar
q}}\over{d{\hat t}}} &=& {{d\sigma^{gq\rightarrow gq}}\over{d{\hat t}}}, \\ 
{{d\sigma^{\bar q \bar q\rightarrow \bar q \bar q}}\over{d{\hat t}}} &=&
{{d\sigma^{qq\rightarrow qq}}\over{d{\hat t}}}.
\end{eqnarray}

We consider a head-on collision ($b=0$) and use the running coupling constant $\alpha_s = {{4 \pi}\over{b_0
\log (Q / \Lambda_{QCD})}}$ where $Q$ is the momentum transfer, $b_0 = 11-{2\over 3}n_f$, and 
$\Lambda_{QCD} = 200$ MeV. 

The production of partons strongly depends on the minimum momentum transfer $Q_0$.
We will set in between $Q_0 = 1.2$ and $2.0$ $GeV/c$ at RHIC energy and $Q_0=2.0$ and $4.0$ $GeV/c$ at LHC energy.
These partons have quite high energy so that they will be called minijets in this study.
Since the nucleons in a nucleus are treated independently, we can write the parton distribution of the nucleus A
as follows;
\begin{eqnarray}
f_{i/A}(x,Q^2) = f_{i/N}(x,Q^2) R_A(x,Q^2),
\end{eqnarray}
where $f_{i/N}(x,Q^2)$ is the parton distribution within a free nucleon and 
$R_A(x,Q^2)$ is the the nucleus ratio function, which is the distribution of a nucleon within the nucleus.

We use the CTEQ4\cite{cteq} or the GRV98\cite{grv98} distribution function for a free nucleon parton distribution 
and the EKS98 parametrization for the ratio function \cite{eks99}. 
We also use the emperical distribution of a nucleus\cite{baier,hirano},
\begin{eqnarray}
x G(x,Q^2) = A \log( {{Q^2 + \Lambda^2} \over \Lambda_0^2} ) x^{-\lambda}{(1-x)^{n}}
\end{eqnarray}
where we set $A = 0.3$, $\Lambda = \Lambda_0 = 0.2 $, $n = 4$ and $\lambda = 0.2 $.

We compare our data with those of Krasnitz et al.\cite{knv03};
\begin{eqnarray}
{{ dN } \over { dy d^2 k_\perp }} &=& {{ \pi R^ 2 } \over { g^ 2} } f(k_\perp  / \Lambda_s )
\end{eqnarray}
where
\begin{eqnarray}
 f &=&  {1 \over g^2} \left\{ 
\begin{array}{l} 
 a_1 [ \exp(\sqrt{k_\perp^2 + m^2} / T_{eff} ) -1 ]^{-1} , ~ k_\perp /\Lambda_s < 1.5  \\ 
 a_2 \Lambda_s ^4 \log( 4 \pi k_\perp / \Lambda_s ) k_\perp^{-4} , ~ k_\perp / \Lambda_s > 1.5  
\end{array} \right. 
\end{eqnarray}
and $a_1 = 0.137, a_2 = 0.0087, m = 0.0358 \Lambda_s , T_{eff} = 0.465 \Lambda_s $\\

\section{Results and Discussions}

Fig. 1 shows the total energy liberated after collision at RHIC energy.
Hirano distribution gives less energy since the distribution does not include quarks and antiquarks. 
GRV and CTEQ function give almost the same results.
The available total energy is about 39.4 TeV for the head-on collision at RHIC.
There should be lots of soft partons and they will take substantial amount of energy.
If we estimate that the total energy liberated
to the minijets is 50-60\% of total available CM energy\cite{brahms}, 
we can set the minimum momentum transfer
to be in-between 1.8 and 2.0 GeV/c, and for 60-70\% liberated energy, $Q_0$ is in-between 1.7 and 1.9 and 
for 70-80\%  about 1.7 GeV/c. These number are larger than those of theoretically predicted, $Q_0^2  = 1 - 2 GeV^2 /c^2 $.
We further notice that the Hirano distribution predictes much lower values than those of GRV and CTEQ distribution 
but gives reasonable theoretical value for the minimum momentum transfer scale, $Q_0$.
\begin{figure}
\includegraphics{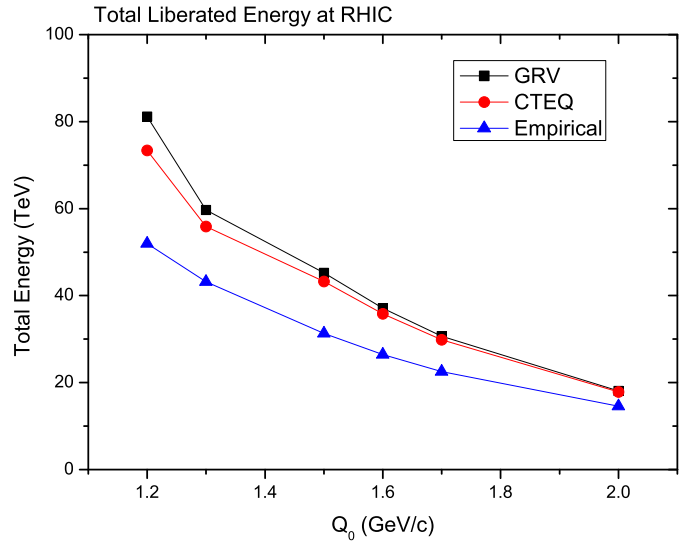}
\caption{Total energy liberated from Au-Au at 200 GeV/pair.} \label{fig1}
\end{figure}

Fig. 2 shows the total energy liberated at LHC which has 1140 TeV available CM energy.
GRV simulation gives more energy deposition than the other two distributions. 
Assuming that the liberated energy into minijets is about 70-80\%,
we can estimate the minimum momentum transfer is 3.5-3.7 GeV/c for CTEQ distribution and
3.7-3.9 GeV/c for GRV model. These numbers are much larger than those of educated theoretical prediction,
$Q_0 ^2 = 2 - 3 GeV^2 /c^2$.
\begin{figure}
\includegraphics{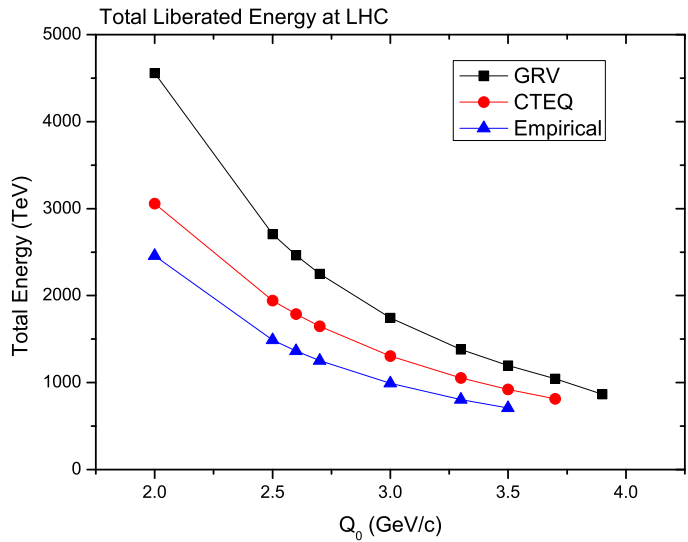}
\caption{Total energy liberated from Pb-Pb at 5.5 TeV/pair.} \label{fig2}
\end{figure}\\

\begin{figure}
\includegraphics{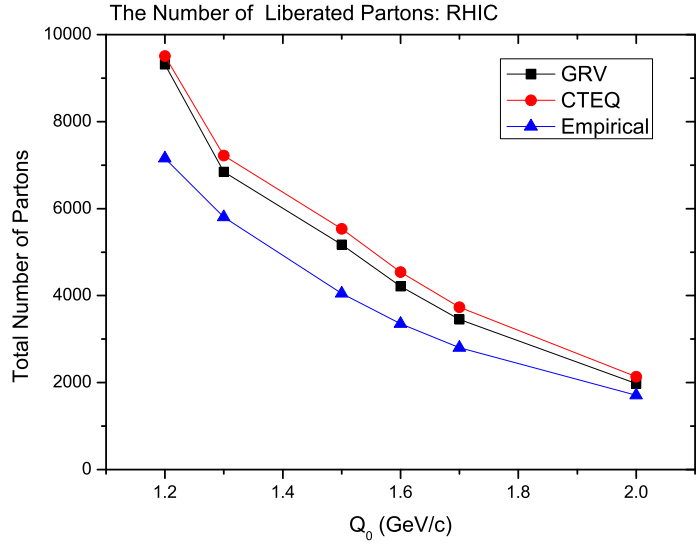}
\caption{Total number of partons produced at RHIC energy.} \label{fig3}
\end{figure}

\begin{figure}
\includegraphics{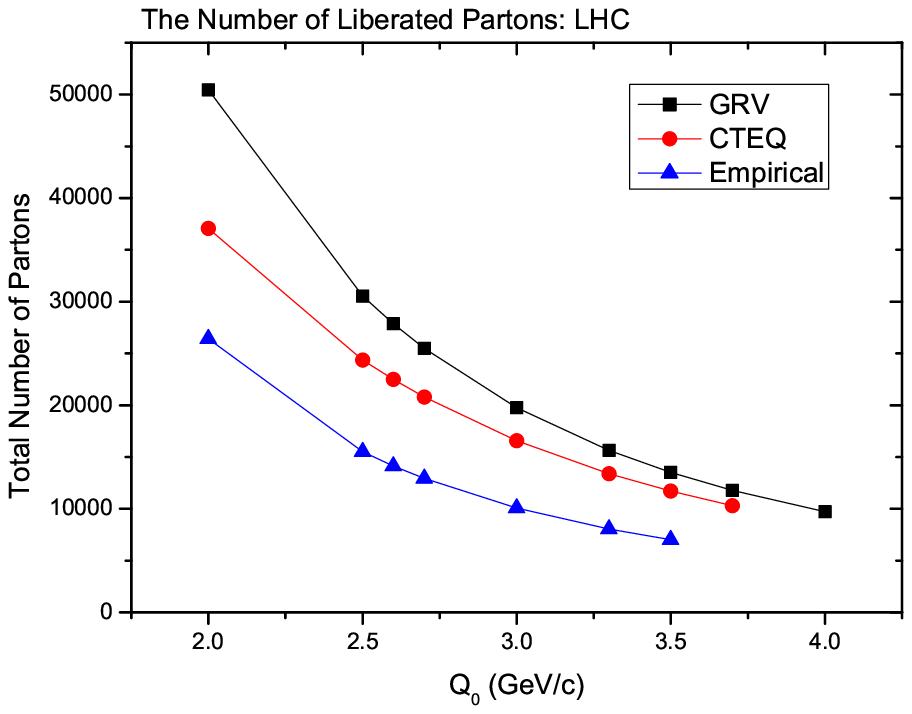}
\caption{Total number of partons produced at LHC energy.} \label{fig4}
\end{figure}

Fig.  \ref{fig3} shows the total number of produced partons at RHIC as a function of $Q_0$ and 
the number of produced partons is about 3700 with CTEQ at $Q_0 = 1.7$ GeV/c, which liberates about 70\% of total energy.
Fig. \ref{fig4} shows the total number of produced partons at LHC as function of $Q_0$. 
The total number of partons with CTEQ is about 11700 at $Q_0 = 3.5 $ GeV/c which liberates about 80\% of CM energy at LHC.
CTEQ predicts more parton production than the others at RHIC but GRV gives far more
partons at LHC, which means GRV distribution have more increasement than the others as $x$ gets smaller.
\\
\\

Table 1 gives the total numbers and the average energy 
in GeV of partons after a Au-Au collision at 200A GeV and a Pb-Pb collision at 5.5A TeV.
\begin{table}
\begin{tabular}{|c|c|c|c|c|} \hline
& \multicolumn{2}{c}{RHIC: $Q_0$ = 1.7 GeV} \vline & 
\multicolumn{2}{c}{LHC: $Q_0$ = 3.5 GeV} \vline 
\\ \hline
 parton  & N &  $\bar{E}$  &  N &  $\bar{E}$  \\ \hline
g & 3062 & 7.01  & 10380 & 66.8  \\ \hline
u & 230 & 13.9 & 356 & 235.7  \\ \hline
d & 255 & 14.5 & 384 & 253.3  \\ \hline
s &  33 &   7.9 & 122 & 79.9  \\ \hline
$\bar{u}$ & 62 & 8.1 & 170 & 90.2  \\ \hline
$\bar{d}$ & 61 & 8.2 & 168 & 92.0  \\ \hline
\end{tabular}
\caption{Initial distribution data: using CTEQ distribution function.}
\end{table}
The table shows that most of partons produced at LHC energy are jets and the liberated valence quarks
have extremely high energy. 30\% of valence quarks are liberated from the colliding nuclei at RHIC
and 32\% at LHC so that substantial amount of baryons are provided into the parton system
from the very beginning. We also note that the strangeness is produced from the primary collisions.
These initial baryon and stageness contents should be imprinted in further evolution of the system.

\begin{figure}
\includegraphics{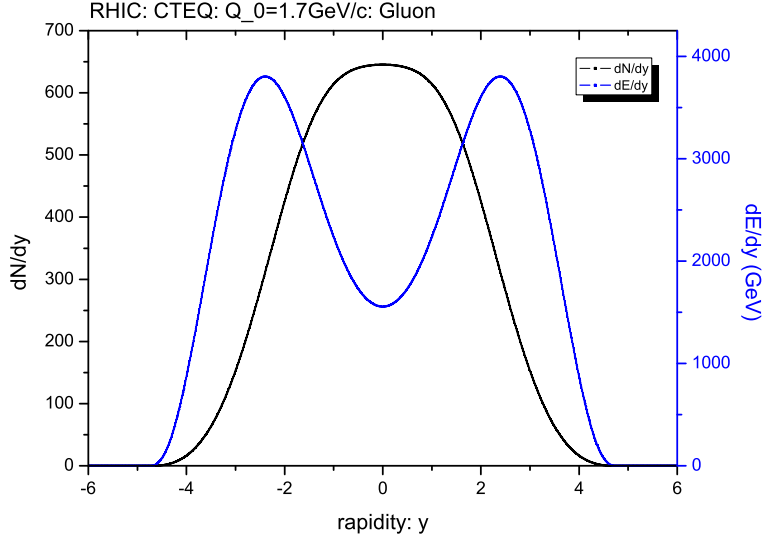}
\caption{Gluon distribution as a function of rapidity at RHIC energy.} \label{fig5}
\end{figure}

\begin{figure}
\includegraphics{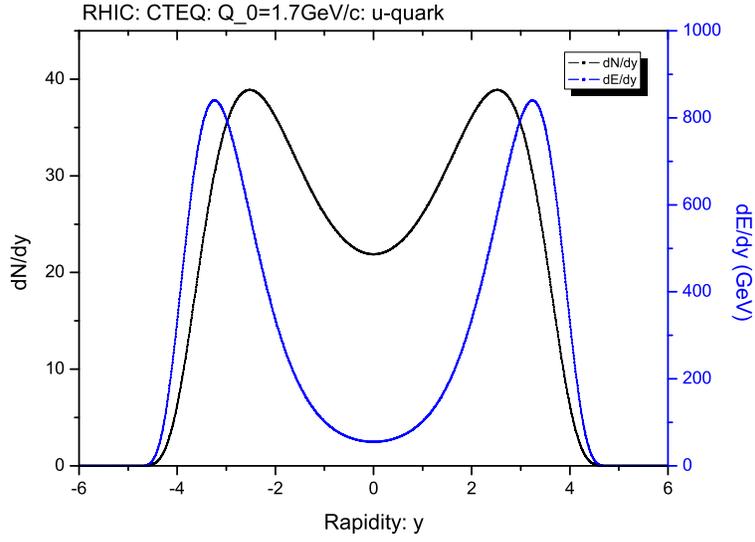}
\caption{u-quark distribution as a function of rapidity at RHIC.} \label{fig6}
\end{figure}

Fig. \ref{fig5} shows the number of gluons produced and energy distribution as a function of
rapidity at RHIC energy.
\\
Even though the gluons are flat and high at central rapidity region, those gluons are strongly oriented
in forward and backward direction energetically. We estimate the $ {{dE}\over{dy}} / {{dN}\over{dy}} $ 
$\approx 2.3$ GeV at central rapidity region.
This average energy of parton at central rapidity region is much lower than that of minijets.
Fig. \ref{fig6} shows the number of u-quarks produced and the energy distribution as a function of
rapidity at RHIC energy. We can see clearly majority of partons move forward or backward in number and energy,
which are mostly valence quarks.
\\

\begin{figure}
\includegraphics{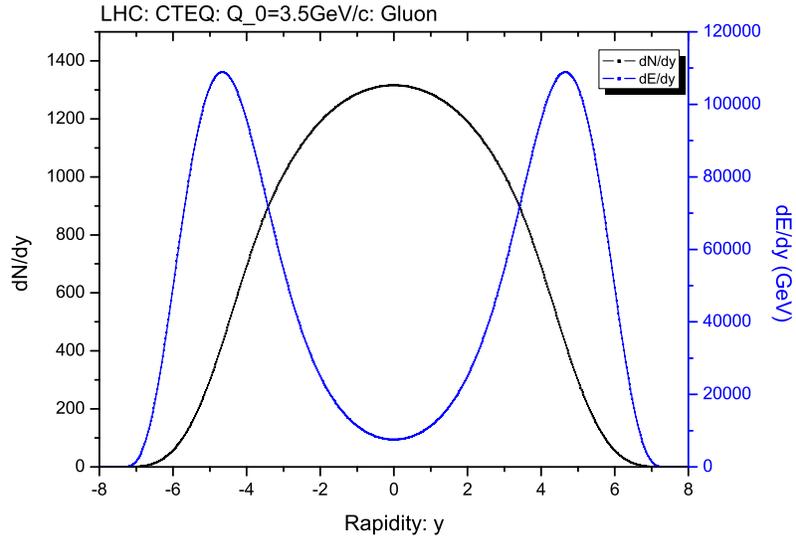}
\caption{Gluon distribution as a function of rapidity at LHC energy.} \label{fig7}
\end{figure}

On the other hand, fig. \ref{fig7} shows the number of gluons produced and the energy distribution as a function of
rapidity at LHC energy. The average energy per gluon is ${{dE}\over{dy}} / {{dN}\over{dy}} $ 
$\approx 5.7$ GeV at central rapidity region. This average energy is more than 10 times smaller compared to those of jets. 
\\

\begin{figure}
\includegraphics{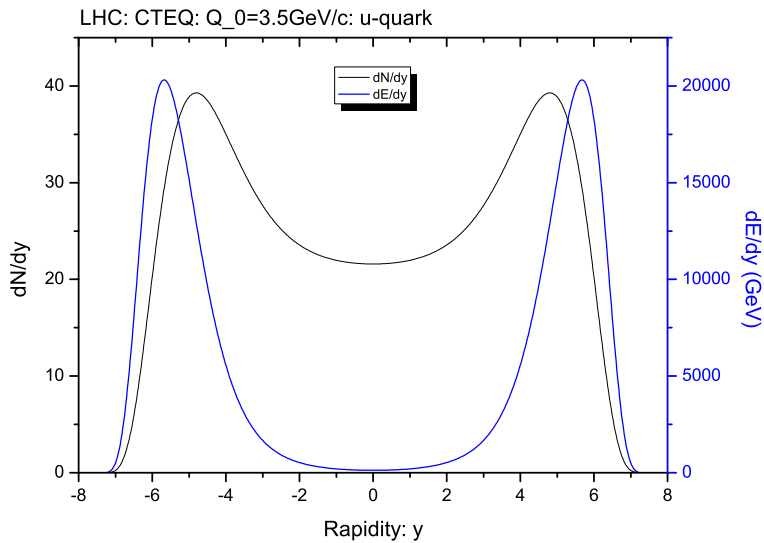}
\caption{u-quark distribution as a function of rapidity at LHC.} \label{fig8}
\end{figure}

Fig. \ref{fig8} shows the number of u-quarks produced and the energy distribution as a function of
rapidity at LHC energy.
\\

\begin{figure}
\includegraphics{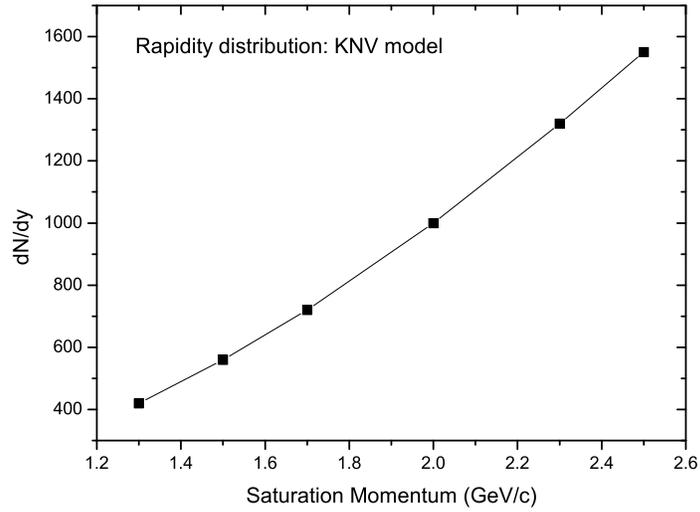}
\caption{Number distribution of KNV distribution.} \label{fig9}
\end{figure}

Fig. \ref{fig9} shows the number density $dN/dy$ at central rapidity region using the KNV function as a function
of the saturation momentum, $Q_s$, where the saturation momentum $Q_s$ is different from the minimum momentum transfer $Q_0$.
Assuming that the distribution is flat inbetween -2.5 and +2.5 and falls off to zero
at $\pm 5$ linearly, the number of parton produced can be obtained by integration over transverse
momentum and rapidity. This number however depends on the saturation momentum, $Q_s$,
which is introduced in CGC scheme and estimated to be 1 GeV/c $<$ $Q_s$ $<$ 2 GeV/c at RHIC energy.
The total number of initial partons is about 5400 for the $Q_s = 1.7$ GeV/c and 4200 for the $Q_s = 1.5$ GeV/c.
We can see that the gluon density at central rapidity region at $Q_0 = 1.7 $ GeV/c is compared to that of KNV results
with $Q_s = 1.7$ GeV/c at RHIC energy and that at $Q_0 = 3.5$ GeV/c compared to $Q_s = 2.3$ GeV/c at LHC energy.
\\

\begin{figure}
\includegraphics{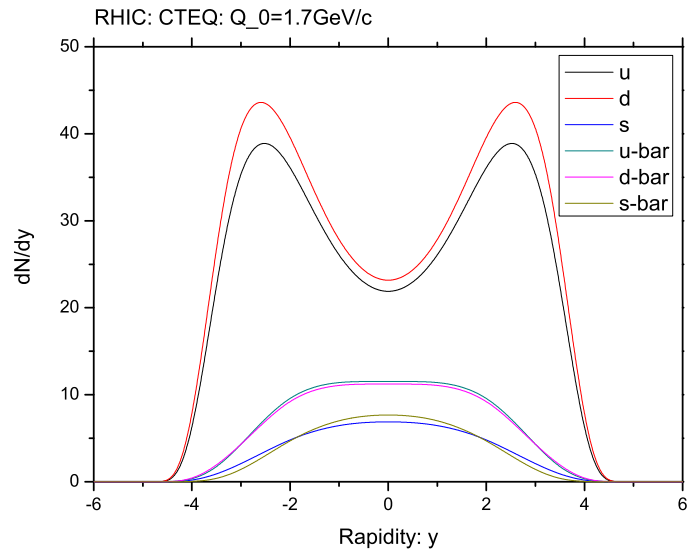}
\caption{Number distribution of quarks and antiquarks.} \label{fig10}
\end{figure}

\begin{figure}
\includegraphics{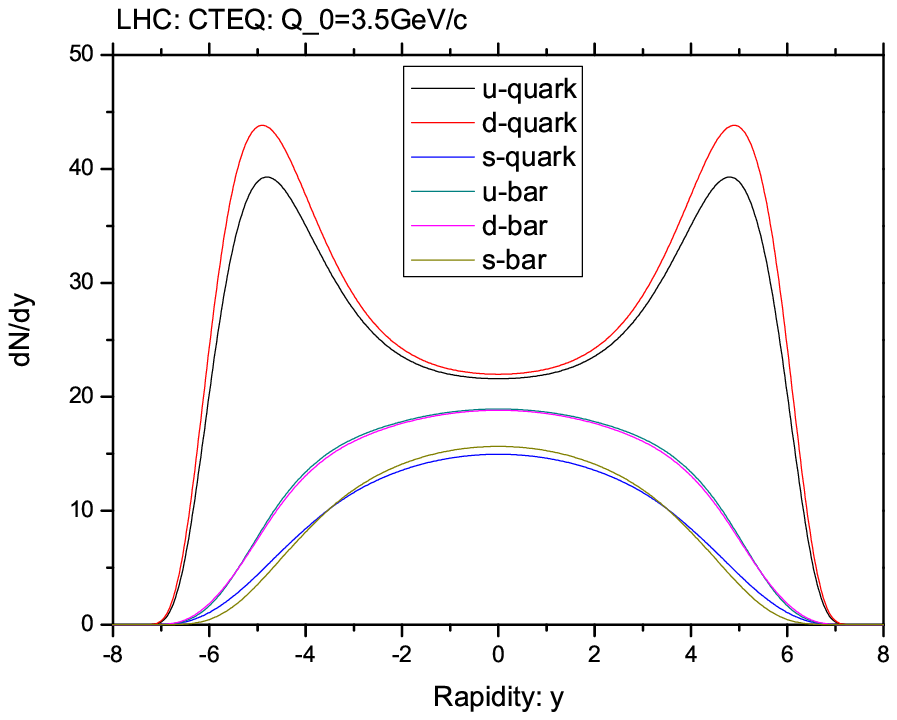}
\caption{Rapidity distribution of quarks and antiquarks at LHC.} \label{fig11}
\end{figure}

Fig. \ref{fig10} and \ref{fig11} show the rapidity distribution of quarks and antiquarks at RHIC and at LHC,
respectively. We have more d-quarks than u-quarks since the Au or Pb have more neutrons than protons.
The difference between a quark and its antiquark is the valence quark.
\\
\\

\begin{figure}
\includegraphics{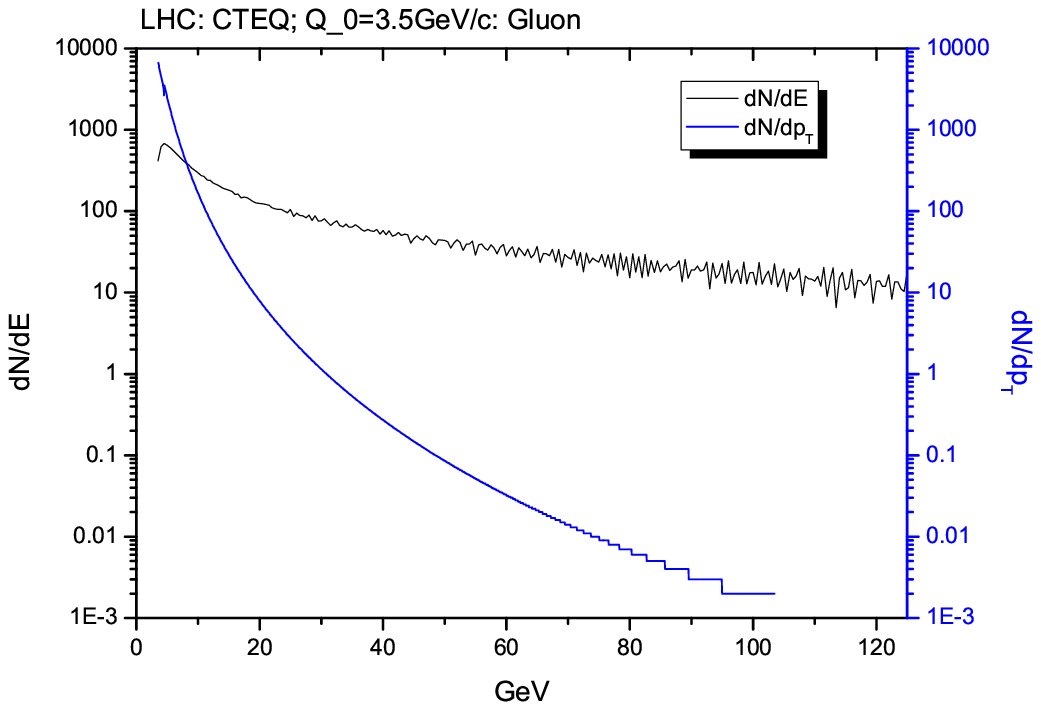}
\caption{Energy and $p_T$ distribution of gluons at LHC.} \label{fig12}
\end{figure}

\begin{figure}
\includegraphics{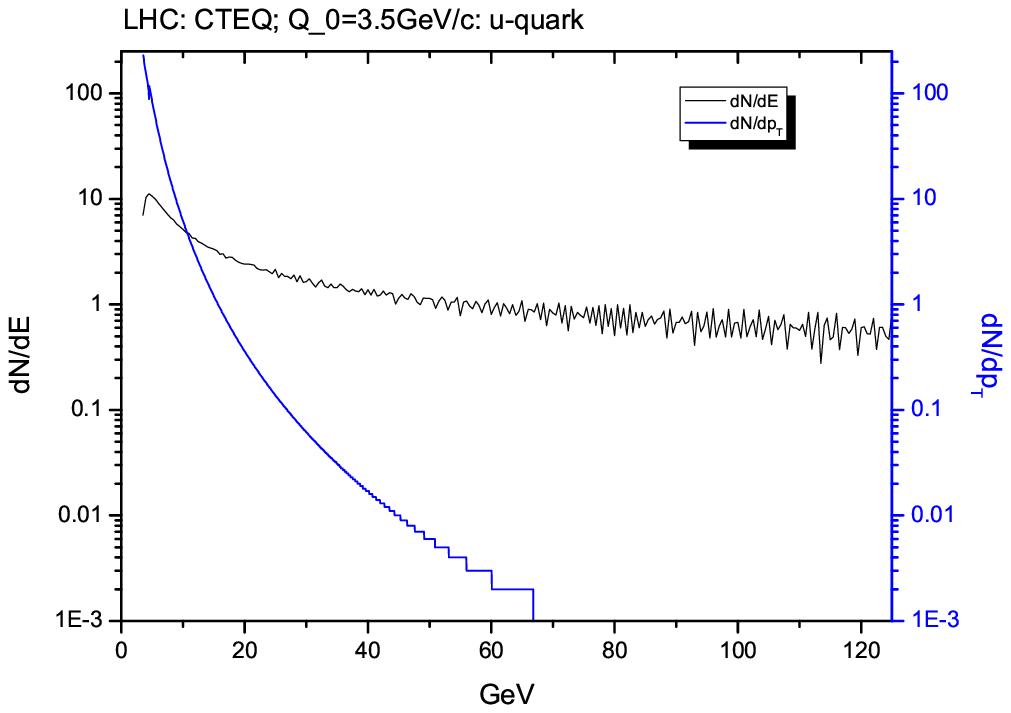}
\caption{Energy and $p_T$ distribution of quarks and antiquarks at LHC.} \label{fig13}
\end{figure}

Fig. \ref{fig12} and \ref{fig13} give the energy and $p_T$ distributions of produced gluons and u-quarks at LHC.
\\
\\

\begin{figure}
\includegraphics{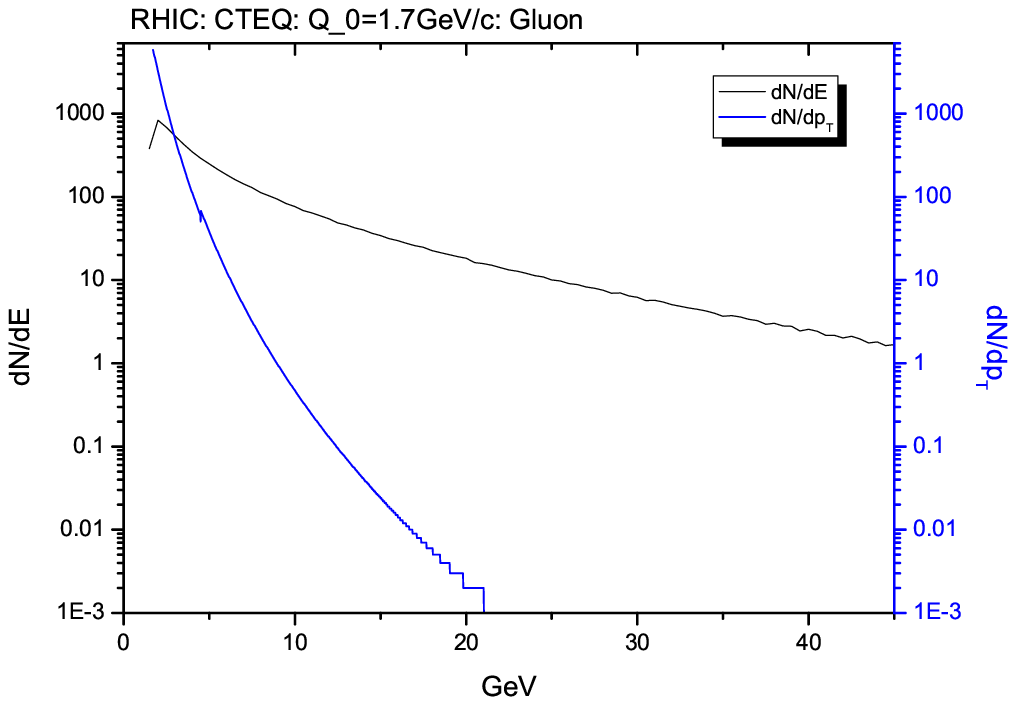}
\caption{Energy and $p_T$ distribution of gluons at RHIC.} \label{fig14}
\end{figure}

\begin{figure}
\includegraphics{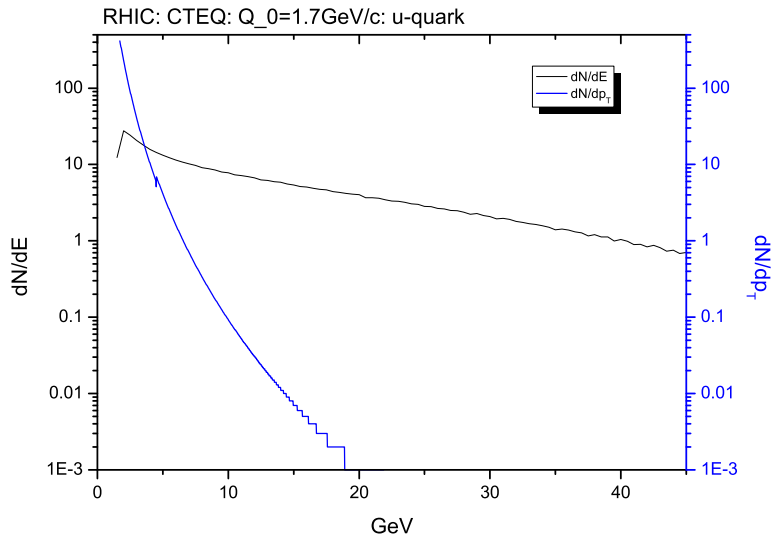}
\caption{Energy and $p_T$ distribution of quarks and antiquarks at RHIC.} \label{fig15}
\end{figure}

Fig. \ref{fig14} and \ref{fig15} give the energy and $p_T$ distributions of produced gluons and u-quarks at RHIC.
\\
\\

\section{Summary and Conclusions}
We showed total energy liberated from heavy ion collisions at RHIC and LHC energy
and estimated the minimum momentum transfer, $Q_0$. We calculated the number of partons
in detail produced from the collisions. 
We presented the rapidity distribution, energy distribution and $p_T$ distribution.

These distributions can be used as initial states to evolve the system. The partonic transport
theory will be ideal to study the evolution of the system \cite{shin02, shin03, bla87, elze, gyu97,gei92} as a function of time.
We  further note that these initial distributions could provide the initial condition for the hydrodynamic
equation of motions.

Acknowledgement: 
This work was supported by the Korea Science and Engineering Foundation(KOSEF) under contract R01-2005-000-10334-0
and Kang Seog Lee is supported by the post-BK21 program.\\

\end{document}